\documentclass[sigconf]{acmart}

\AtBeginDocument{%
  \providecommand\BibTeX{{%
    \normalfont B\kern-0.5em{\scshape i\kern-0.25em b}\kern-0.8em\TeX}}}
\usepackage{array,booktabs,ragged2e}

\usepackage{algorithm}
\usepackage{enumitem}
\usepackage{multirow}
\usepackage{tabularx}
\usepackage{subfigure}
\usepackage{fontenc}
\usepackage{float}
\usepackage{siunitx}

\usepackage[noend]{algpseudocode}

\copyrightyear{2020} 
\acmYear{2020} 
\setcopyright{acmcopyright}\acmConference[SIGIR '20]{Proceedings of the 43rd International ACM SIGIR Conference on Research and Development in Information Retrieval}{July 25--30, 2020}{Virtual Event, China}
\acmBooktitle{Proceedings of the 43rd International ACM SIGIR Conference on Research and Development in Information Retrieval (SIGIR '20), July 25--30, 2020, Virtual Event, China}
\acmPrice{15.00}
\acmDOI{10.1145/3397271.3401116}
\acmISBN{978-1-4503-8016-4/20/07}

\usepackage{amsmath,amsfonts,bm}

\def\eqref#1{equation~\ref{#1}}

\def\1{\bm{1}}

\def\vb{{\bm{b}}}

\def\ve{{\bm{e}}}

\def\vh{{\bm{h}}}

\def\vx{{\bm{x}}}

\def\mD{{\bm{D}}}
\def\mE{{\bm{E}}}

\def\mR{{\bm{R}}}

\def\mW{{\bm{W}}}
\def\mX{{\bm{X}}}

\DeclareMathAlphabet{\mathsfit}{\encodingdefault}{\sfdefault}{m}{sl}
\SetMathAlphabet{\mathsfit}{bold}{\encodingdefault}{\sfdefault}{bx}{n}

\def\gG{{\mathcal{G}}}

\def\gI{{\mathcal{I}}}

\def\gN{{\mathcal{N}}}

\def\gU{{\mathcal{U}}}

\def\sR{{\mathbb{R}}}

\def\sZ{{\mathbb{Z}}}

\newcommand{\E}{\mathbb{E}}

\newcommand{\name}{GroupIM}

\begin{document}
\fancyhead{}

\title{GroupIM: A Mutual Information Maximization Framework for Neural Group Recommendation}

\author{Aravind Sankar$^*$, Yanhong Wu$^\dagger$, Yuhang Wu$^\dagger$, Wei Zhang$^\dagger$, Hao Yang$^\dagger$, Hari Sundaram$^*$}
\affiliation{
  \footnotemark[1]\institution{University of Illinois at Urbana-Champaign, IL, USA}
  \footnotemark[2]\institution{Visa Research, Palo Alto, CA, USA}
  \footnotemark[1]\{asankar3, hs1\}@illinois.edu \hspace{10pt} \footnotemark[2]\{yanwu, yuhawu, wzhan, haoyang\}@visa.com
}

\renewcommand{\authors}{Aravind Sankar, Yanhong Wu, Yuhang Wu, Wei Zhang, Hao Yang, Hari Sundaram}

\renewcommand{\shortauthors}{Trovato and Tobin, et al.}
\copyrightyear{2020}
\acmYear{2020}
\setcopyright{acmcopyright}

\begin{abstract}

We study the problem of making item recommendations to ephemeral groups, which comprise users with limited or no historical activities together.
Existing studies target persistent groups with substantial activity history, while ephemeral groups lack historical interactions.
To overcome group interaction sparsity, we propose data-driven \textit{regularization} strategies to exploit both the \textit{preference covariance} amongst users who are in the same group, as well as the \textit{contextual relevance} of users' individual preferences to each group.

We make two contributions.
First, we present a recommender architecture-agnostic framework GroupIM that can integrate arbitrary neural preference encoders and aggregators for ephemeral group recommendation.
Second, we regularize the user-group latent space to overcome group interaction sparsity by: maximizing \textit{mutual information} between representations of groups and group members; and dynamically prioritizing the preferences of highly informative members through contextual preference weighting.
Our experimental results on several real-world datasets indicate significant performance improvements (31-62\% relative NDCG@20) over state-of-the-art group recommendation techniques.

\end{abstract}


\newcommand{\yh}[1]{{\color{blue}{<#1>}}}

\setlist[itemize]{leftmargin=*, topsep=1pt, itemsep=1pt, parsep=0.5pt, partopsep=0pt}

\setlist[enumerate]{leftmargin=*, topsep=0pt, itemsep=0pt, parsep=0pt, partopsep=0pt}

\setlength{\abovedisplayskip}{0.05cm}
\setlength{\belowdisplayskip}{0.05cm}

\setlength{\floatsep}{0.1cm}

\setlength{\textfloatsep}{0.1cm}

\setlength{\abovecaptionskip}{0.05cm}

\setlength{\belowcaptionskip}{0.05cm}

\setlength{\dbltextfloatsep}{0.05cm}

\setlength{\intextsep}{0.1cm}

\theoremstyle{mystyle}

\newtheorem{define}{Definition}

\maketitle

\section{Introduction}

We address the problem of recommending items to \textit{ephemeral}
groups, which comprise users who purchase \textit{very few (or no)} items together~\cite{ephemeral}.
The problem is ubiquitous, and appears in a variety of familiar contexts, \textit{e.g.}, dining with strangers, watching movies with new friends, and attending social events.
We illustrate key challenges with an example: 
Alice (who loves Mexican food) is taking a visitor Bob (who loves Italian food) to lunch along with her colleagues, 
where will they go to lunch? 
There are three things to note here:
first, the group is \textit{ephemeral}, since there is \textit{no}  historical interaction observed for this group.
Second, 
\textit{individual preferences may depend on other group members}.
In this case, the group may go to a fine-dining Italian restaurant. 
However, when Alice is with other friends, they may go to
Mexican restaurants.
Third, groups comprise users with \textit{diverse individual preferences}, and thus the group recommender needs to be cognizant of individual preferences.

Prior work primarily target persistent groups which refer to fixed, stable groups where members have interacted with numerous items \textit{as a group} (\textit{e.g.}, families watching movies).
They mainly
fall into two categories: heuristic pre-defined aggregation (\textit{e.g.}, least misery~\cite{least_misery}) %
that disregards group interactions; 
data-driven strategies such as
probabilistic models~\cite{com, crowdrec} and neural preference aggregators~\cite{agree, agr}.
A key weakness is that these methods either ignore individual user activities~\cite{agr, wang2019group} or assume that users have the same likelihood to follow individual and collective preferences, across different groups~\cite{com, crowdrec, agree}.
Lack of expressivity to distinguish the role of individual preferences across groups results in degenerate solutions for sparse ephemeral groups. %
A few methods exploit external side information in the form of a social network~\cite{socialgroup,SoAGREE}, user personality traits and demographics~\cite{delic2018observational}, for group decision making.
However, side information may often be unavailable.

We train robust ephemeral group recommenders without resorting to any extra side information.
Two observations help: first, while groups are ephemeral, group members may have rich individual interaction histories; this can alleviate group interaction sparsity.
Second, since groups are ephemeral with sparse training interactions, base group recommenders need reliable guidance to learn informative (non-degenerate) group representations,
but the guidance needs to be data-driven, rather than a heuristic.

To overcome group interaction sparsity, our key technical insight is to regularize the latent space of user and group representations in a manner that exploits the \textit{preference covariance} amongst individuals who are in the same group, as well as to incorporate
the \textit{contextual} relevance of users' personal preferences to each group.

Thus, we propose two data-driven regularization strategies. \textit{First}, we contrastively regularize the user-group latent space
to capture social user associations and distinctions across groups.
We achieve this by 
maximizing \textit{mutual information} (MI) between representations of groups and group members,
which encourages group representations to encode shared group member preferences while regularizing user representations to capture their social associations.
\textit{Second}, we contextually identify \textit{informative} group members and regularize the corresponding group representation to reflect their personal preferences.
We introduce a novel regularization objective that \textit{contextually} weights users' personal preferences in each group, in proportion to their user-group MI.
\textit{Group-adaptive} preference weighting precludes
degenerate solutions that arise during static regularization over ephemeral groups with sparse activities.
We summarize our key contributions below:

\begin{itemize}[leftmargin=*]
    \item \textbf{Architecture-agnostic Framework%
    }: 
    To the best of our knowledge, \textbf{Group} \textbf{I}nformation \textbf{M}aximization (\name) is the first recommender architecture-agnostic framework for group recommendation. Unlike prior work~\cite{agr, agree} that design customized preference aggregators, \name~can integrate arbitrary neural preference encoders and aggregators. We show state-of-the-art results with simple efficient aggregators (such as meanpool) that are contrastively regularized within our framework. 
The effectiveness of meanpool signifies substantially \textit{reduced inference costs} without loss in model expressivity. Thus,~\name~facilitates straightforward enhancements to base neural recommenders. %
    \item \textbf{Group-adaptive Preference Prioritization}: We learn robust estimates of group-specific member relevance. In contrast, prior work incorporate personal preferences through static regularization~\cite{com, crowdrec, agree}. 
    We use \textit{Mutual Information} to dynamically learn user and group representations that capture 
    preference covariance across individuals in the same group; and prioritize the preferences of highly relevant members through group-adaptive preference weighting; thus effectively overcoming group interaction sparsity in ephemeral groups.
    An ablation study confirms the superiority of our MI based 
    regularizers over static alternatives.
    \item \textbf{Robust Experimental Results}: 
    Our experimental results indicate significant performance gains for GroupIM over state-of-the-art group recommenders on four publicly available datasets (relative gains of 31-62\% NDCG@20 and 3-28\% Recall@20). Significantly, GroupIM achieves stronger gains for: groups of larger sizes; and groups with diverse member preferences.
\end{itemize}

We organize the rest of the paper as follows.
In Section 3, we formally define the problem, introduce a base group recommender unifying existing neural methods, and discuss its limitations.
We describe our proposed framework~\name~in Section 4, present experimental results in Section 5, finally concluding in Section 6.

\section{Related Work}
\textbf{Group Recommendation:} This line of work can be divided into two categories based on group types:
\textit{persistent} and \textit{ephemeral}. 
Persistent groups have stable members with rich activity history together, while ephemeral groups comprise users who interact with very few items together~\cite{ephemeral}.
A common approach is to consider persistent groups as virtual users~\cite{dlgr}, thus, personalized recommenders can be directly applied.
However, such methods cannot handle ephemeral groups with sparse
interactions.
We focus on the more challenging scenario---\textit{recommendations to ephemeral groups}.

Prior work either aggregate recommendation results (or item scores) for each member, or aggregate individual member preferences, towards group predictions.
They fall into two classes: \textit{score (or late)} aggregation~\cite{least_misery} and \textit{preference (or early)} aggregation~\cite{com}.

Popular \textit{score aggregation} strategies include least misery~\cite{least_misery}, average~\cite{average}, maximum satisfaction~\cite{max_satisfaction}, and relevance and disagreement~\cite{rd}.
However, these are hand-crafted heuristics that overlook real-world group interactions.
~\citet{least_misery} compare different strategies to conclude that there is no clear winner, and their relative effectiveness depends on group size and group coherence.

Early \textit{preference aggregation} strategies~\cite{merging} generate recommendations by constructing a group profile that combines the profiles (raw item histories) of group members.
Recent methods adopt a \textit{model-based} perspective to learn data-driven models.
\textit{Probabilistic} methods~\cite{pit,com,crowdrec} model the group generative process by considering both the personal preferences and relative influence of members, to differentiate their contributions towards group decisions.
However, a key weakness is their assumption that users have the same likelihood to follow individual and collective preferences, across different groups.
\textit{Neural} methods explore attention mechanisms~\cite{attention} to learn data-driven preference aggregators~\cite{agree, agr, wang2019group}.
MoSAN~\cite{agr} models group interactions via sub-attention networks; however, MoSAN operates on persistent groups while ignoring users' personal activities.
AGREE~\cite{agree} employs attentional networks for joint training over individual and group interactions; yet, the extent of regularization applied on each user (based on personal activities) is the same across groups, which results in degenerate solutions when applied to ephemeral groups with sparse activities.

An alternative  approach to tackle interaction sparsity is to exploit external side information, \textit{e.g.}, social network of users~\cite{SoAGREE,socialgroup, infvae, advsorec}, personality traits~\cite{zheng2018identifying}, demographics~\cite{delic2018observational}, and interpersonal relationships~\cite{delic2018use, gartrell2010enhancing}.
In contrast, our setting is conservative and does not include extra side information: we know only user and item ids, and item implicit feedback.
We address interaction sparsity through novel data-driven regularization and training strategies~\cite{longtail}.
Our goal is to enable a wide spectrum of neural group recommenders to seamlessly integrate suitable preference encoders and aggregators.

\textbf{Mutual Information:}
Recent neural MI estimation methods~\cite{mine} leverage the InfoMax~\cite{infomax} principle for representation learning.
They exploit the \textit{structure} of the input data (\textit{e.g.}, spatial locality in images) via MI maximization objectives, to improve representational quality.
Recent advances employ auto-regressive models~\cite{cpc} and aggregation functions~\cite{dim, dgi, yeh2019qainfomax} with noise-contrastive loss functions to preserve MI between structurally related inputs.

We leverage the InfoMax principle to exploit the preference covariance \textit{structure} shared amongst group members.
A key novelty of our approach is MI-guided weighting to regularize group embeddings with the personal preferences of highly relevant members.

\section{Preliminaries}
In this section, we first formally define the ephemeral group recommendation problem. 
Then, we present a base neural group recommender $\mR$ that unifies existing neural methods into a general framework.
Finally, we analyze the key shortcomings of $\mR$ to 
discuss motivations for maximizing user-group mutual information.

\subsection{Problem Definition}
We consider the implicit feedback setting (only visits, no explicit ratings) with a user set $\gU$, an item set $\gI$, a group set $\gG$, a binary $|\gU| \times |\gI|$ user-item interaction matrix $\mX_U$,
and a binary $|\gG| \times |\gI|$ group-item interaction matrix $\mX_G$. We denote $\vx_u$, $\vx_g$ as the corresponding rows for user $u$ and group $g$ in $\mX_U$ and $\mX_G$, with $|\vx_u|$, $|\vx_g|$ indicating their respective number of interacted items.

An \textit{ephemeral group} $g \in \gG$ comprises a set of $|g|$ users $u^g = \{u^g_1, \dots, u^g_{|g|} \} \subset \gU$ with sparse historical interactions $\vx_g$.

\textbf{Ephemeral Group Recommendation}: 
We evaluate group recommendation on \textit{strict ephemeral groups}, which have never interacted together before during training.
Given a strict ephemeral group $g$ during testing, 
our goal is to generate a ranked list over the item set $\gI$ relevant to users in $u^g$, \textit{i.e.}, learn a 
function $f_G : P(\gU) \times \gI \mapsto \sR$
that maps an ephemeral group and an item to a relevance score, where $P(\gU)$ is the power set of $\gU$.

\subsection{Base Neural Group Recommender}
\label{sec:base_recommender}
Several neural group recommenders 
have achieved impressive results~\cite{agree,agr}.
Despite their diversity in modeling group interactions, we remark that state-of-the-art neural methods share a clear model structure: we present a base group recommender $\mR$ that includes three modules:
a preference encoder; a preference aggregator; and a joint user and group interaction loss. Unifying these neural group recommenders within a single framework 
facilitates deeper analysis into their shortcomings in addressing ephemeral groups.

The base group recommender $\mR$ first computes user representations $\mE \in \sR^{|\gU| \times D}$ from user-item interactions $\mX_U$ using a preference encoder $f_{\textsc{enc}} (\cdot)$, followed by applying a neural preference aggregator $f_{\textsc{agg}} (\cdot)$ to compute the group representation $\ve_g$ for group $g$. 
Finally, the group representation $\ve_g$ is jointly trained over the group $\mX_G$ and user $\mX_U$ interactions,
to make group recommendations.

\subsubsection{\textbf{User Preference Representations.}}
User embeddings $\mE$ constitute a latent representation of their personal preferences, indicated in the interaction matrix $\mX_U$.
Since latent-factor collaborative filtering methods adopt a variety of strategies (such as matrix factorization, autoencoders, etc.) to learn user embeddings $\mE$, we define the preference encoder $f_{\textsc{enc}} : |\gU| \times \sZ_2^{|\gI|} \mapsto \sR^{D} $ with two inputs: user $u$ and associated binary personal preference vector $\vx_u$.

\begin{equation}
\ve_u =  f_{\textsc{enc}} (u, \vx_u) %
\;  \forall u \in \gU
\label{eqn:user_rep}
\end{equation}

We can augment $\ve_u$ with additional inputs, including contextual attributes, item relationships, etc. via customized encoders~\cite{survey}.

\subsubsection{\textbf{Group Preference Aggregation.}}
A preference aggregator models the interactions among group members to compute an aggregate representation $\ve_g \in \sR^D$ for ephemeral group $g \in \gG$.
Since groups are sets of users with no inherent ordering, we consider the class of permutation-invariant functions (such as summation or pooling operations) on sets~\cite{deep_sets}.
Specifically, $f_{\textsc{agg}} (\cdot)$ is permutation-invariant to the order of group member embeddings $\{ e_{u_1}, \dots, e_{u_{|g|}} \}$.
We compute %
$\ve_g$ %
using an arbitrary preference aggregator $f_{\textsc{agg}} (\cdot)$ as:
\begin{equation}
    \ve_g = f_{\textsc{agg}} (\{ \ve_u : u \in u^g \}) \; \forall g \in \gG 
    \label{eqn:group_rep}
\end{equation}

\subsubsection{\textbf{Joint User and Group Loss.}}
The group representation $\ve_g$ is trained over the group-item interactions $\mX_G$ with group-loss $L_G$. 
The framework supports different recommendation objectives, including pairwise~\cite{bpr} and pointwise~\cite{ncf} ranking losses.
Here, we use a multinomial likelihood formulation owing to its impressive results in user-based neural collaborative filtering~\cite{vae-cf}. The group representation $\ve_g$ is transformed by a fully connected layer and normalized by a softmax function to produce a probability vector $\pi (\ve_g)$ over $\gI$. 
The loss measures the KL-divergence between the normalized purchase history $\vx_{g}/|x_g|$ ($\vx_g$ indicates items interacted by group $g$) and predicted item probabilities $\pi (\ve_g)$, given by:

\begin{equation}
 \hspace{-5pt} L_G = - \sum\limits_{g \in \gG} \frac{1}{|\vx_g|} \sum\limits_{i \in \gI} x_{gi} \log \pi_i(\ve_g) ; \hspace{5pt} \pi(\ve_g)= \text{softmax}(\mW_I \ve_g) 
 \label{eqn:group_loss}
\end{equation}

Next, we define the user-loss $L_U$ that regularizes the user representations $\mE$ with user-item interactions $\mX_U$, thus facilitating joint training 
with shared encoder $f_{\textsc{enc}} (\cdot)$ and predictor ($\mW_I$) layers~\cite{agree}.
We use a similar multinomial likelihood-based formation, given by:
\begin{equation}
    L_U = - \sum\limits_{u \in \gU} \frac{1}{|\vx_u|} \sum\limits_{i \in \gI} x_{ui} \log \pi_i(\ve_u) ; \hspace{5pt} L_R = L_G + \lambda L_U
    \label{eqn:user_loss}
\end{equation}
where $L_R$ denotes the overall loss of the base recommender $\mR$ with balancing hyper-parameter $\lambda$.
AGREE~\cite{agree} trains an attentional aggregator with pairwise regression loss over both $\mX_U$ and $\mX_G$, while MoSAN~\cite{agr} trains a collection of sub-attentional aggregators with bayesian personalized ranking~\cite{bpr} loss on just $\mX_G$.
Thus, state-of-the-art neural methods AGREE~\cite{agree} and MoSAN~\cite{agr} are specific instances of the framework described by base recommender $\mR$.

\subsection{Motivation}
\label{sec:motivations}

To address ephemeral groups, we focus on \textit{regularization} strategies that are \textit{independent} of the base recommender $\mR$. 
With the rapid advances in neural methods, we envision future enhancements in neural architectures for user representations and group preference aggregation.
Since ephemeral groups by definition purchase very few items together, base recommenders suffer from inadequate training data in group interactions.
Here, the group embedding $\ve_g$ receives back-propagation signals from sparse interacted items in $\vx_g$, thus lacking evidence to reliably estimate the role of each member.
To address group interaction sparsity towards robust ephemeral group recommendation, we propose two 
data-driven regularization strategies that are independent of the base recommendation mechanisms to generate individual and group representations.

\vspace{-4pt}
\subsubsection{\textbf{Contrastive Representation Learning}}

We note that users' preferences are \textit{group-dependent}; and users occurring together in groups typically exhibit \textit{covarying} preferences (\textit{e.g.}, shared cuisine tastes).
Thus, group activities reveal \textit{distinctions} across groups (\textit{e.g.}, close friends versus colleagues) and latent \textit{user associations} (\textit{e.g.}, co-occurrence of users in similar groups), that are not evident when the base recommender $\mR$ only predicts sparse group interactions.

We \textit{contrast} the preference representations of group members against those of non-member users with similar item histories, to effectively regularize the latent space of user and group representations.
This promotes the representations to encode 
latent discriminative characteristics shared by group members, that are not discernible from their limited interacted items in $\mX_G$.

\vspace{-4pt}
\subsubsection{\textbf{Group-adaptive Preference Prioritization.}}
To overcome group interaction sparsity, we critically remark that while groups are ephemeral with sparse interactions, the group members have comparatively richer individual interaction histories. Thus, we propose to \textit{selectively} exploit the personal preferences of group members to enhance the quality of group representations.

The user-loss $L_U$ (equation~\ref{eqn:user_loss}) in base recommender $\mR$ attempts to regularize user embeddings $\mE$ based on their individual activities $\mX_U$.
A key weakness is that
$L_U$ forces $\ve_u$ to \textit{uniformly} predict preferences $\vx_u$ across all groups containing user $u$.
Since groups interact with items differently than individual members,
inaccurately utilizing $\mX_U$ can become counter-productive.
Fixed regularization results in degenerate models that either over-fit or are over-regularized, due to lack
of flexibility in adapting preferences per group.

To overcome group interaction sparsity, 
we \textit{contextually} identify members that are highly \textit{relevant} to the group and regularize the group representation to reflect their personal preferences.
To measure contextual relevance, we introduce group-specific relevance weights $w (u, g)$ for each user $u$ where $w(\cdot)$ is a learned weighting function of both user and group representations.
This enhances the expressive power of the recommender, thus effectively alleviating the challenges imposed by group interaction sparsity.

In this section, we defined ephemeral group recommendation, and 
presented a base group recommender architecture with three modules: user representations, group preference aggregation, and joint loss functions.
Finally, 
we motivated the need to:
contrastively regularize the user-group space to capture member associations and group distinctions; and learn group-specific weights $w(u,g)$ to regularize group representations with individual user preferences.

\section{\textsc{GroupIM} Framework}

\label{sec:groupim}
In this section, we first
motivate mutual information towards achieving our two proposed regularization strategies, followed by a detailed description of our proposed framework~\textsc{\name}.

\subsection{\textbf{Mutual Information Maximization.}}

\begin{figure*}[t]
    \centering
    \includegraphics[width=0.97\linewidth]{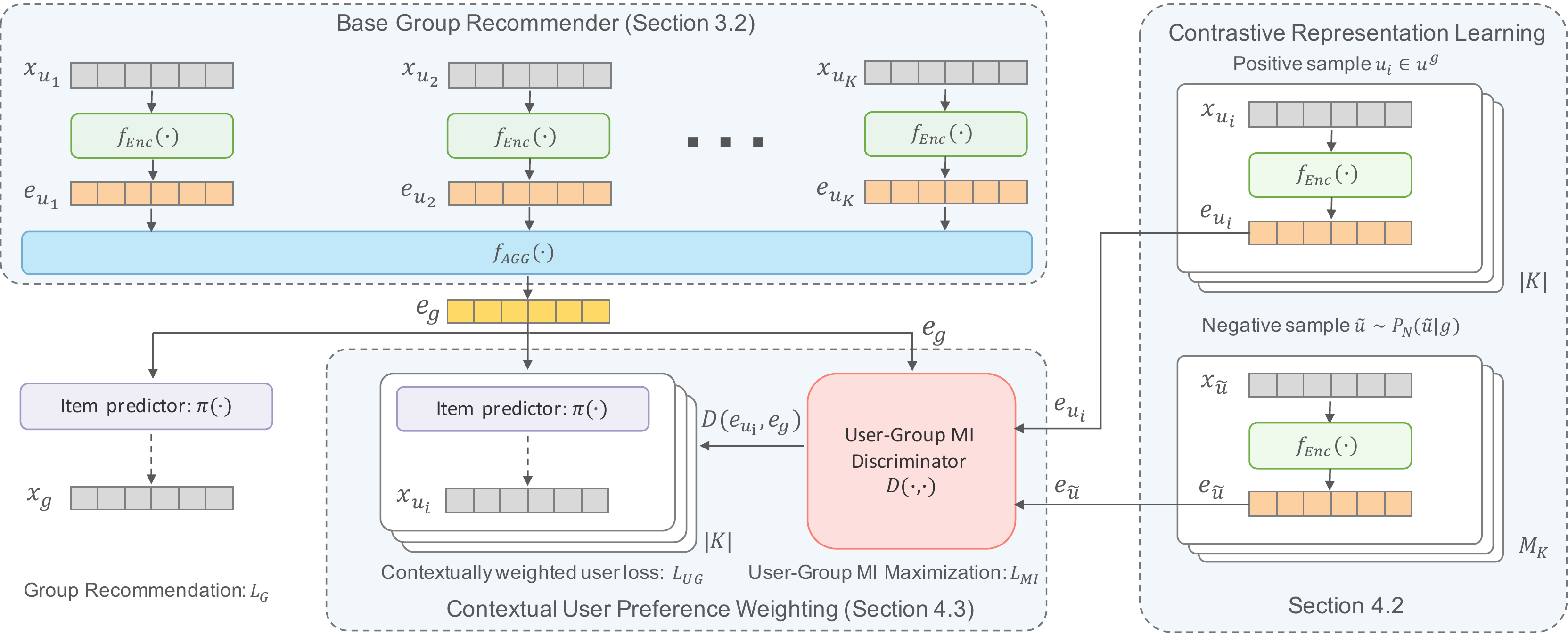}
    \caption{Neural architecture diagram of~\name~depicting model components and loss terms appearing in Equation~\ref{eqn:model_objective}.}
    \vspace{6pt}
    \label{fig:base_grec}
\end{figure*}

We introduce our user-group mutual information maximization approach through an illustration.
We extend the introductory example to illustrate how to regularize Alice's latent representation based on her interactions in two different groups.
Consider Alice who first goes out for lunch to an Italian restaurant with a visitor Bob, and later dines at a Mexican restaurant with her friend Charlie.

First, Alice plays different roles across the two groups (\textit{i.e.}, stronger influence among friends than with Bob) due to the differences in group context (visitors versus friends). Thus, we require a measure to quantify the contextual \textit{informativeness} of user $u$ in group $g$.

Second, we require the embedding of Alice to capture association with both visitor Bob and friend Charlie, yet express variations in her group activities.
Thus, it is necessary to not only \textit{differentiate} the role of Alice across groups, but also compute appropriate representations that make her presence in each group more \textit{coherent}.

To achieve these two goals at once, we maximize \textit{user-group mutual information} (MI) to regularize the latent space of user and group representations, and set group-specific relevance weights $w(u,g)$ in proportion to their estimated MI scores.
User-group MI measures the contextual informativeness of a member $u$ towards the group decision through the reduction in group decision uncertainty when user $u$ is included in group $g$.
Unlike correlation measures that quantify monotonic linear associations, mutual information captures complex non-linear statistical relationships between covarying random variables. Our proposed MI maximization strategy enables us to achieve our two-fold motivation (Section~\ref{sec:motivations}): 

\begin{itemize}
    \item \textbf{Altering Latent Representation Geometry}: %
    Maximizing user-group MI 
    encourages the group embedding
    $\ve_g$ to encode preference covariance across group members, and regularizes the user embeddings $\mE$ to capture social associations in group interactions.
    \item 
    \textbf{Group-specific User Relevance}: 
    By quantifying $w(u,g)$ through
    user-group mutual information, 
    we accurately capture the extent of \textit{informativeness} for user $u$ in group $g$, thus guiding group-adaptive personal preference prioritization.
\end{itemize}

\subsection{\textbf{User-Group MI Maximization.}}
\label{sec:user-group-im}
 Neural MI estimation~\cite{mine} has demonstrated feasibility to maximize MI by training a classifier $\mD$ (\textit{a.k.a}, \textit{discriminator} network) to accurately separate \textit{positive} samples drawn from their joint distribution from \textit{negative} samples drawn from the product of marginals.

We maximize user-group MI between group member representations $\{ \ve_u : u \in u^g \}$ and group representation $\ve_g$ (computed in equations~\ref{eqn:user_rep} and~\ref{eqn:group_rep} respectively).
We train
a contrastive \textit{discriminator} network 
$\mD: \sR^D \times \sR^D \mapsto \sR^{+}$, where $\mD(\ve_u, \ve_g)$ represents the probability score assigned to this user-group pair (higher scores for users who are members of group $g$).
The positive samples $(\ve_u, \ve_g)$ for $\mD$ are the preference representations of $(u,g)$ pairs such that $u \in u^g$, and negative samples are derived by pairing $\ve_g$ with the representations of non-member users sampled from a negative sampling distribution $P_{\gN} (u | g)$. 
The discriminator $\mD$ is trained on a noise-contrastive type objective with a binary cross-entropy (BCE) loss between samples from the joint (positive pairs), and the product of marginals (negative pairs), resulting in the following objective:

\begin{align}
    \hspace{-0.9em} L_{MI} = - \frac{1}{|\gG|} %
    \sum\limits_{g \in \gG} \frac{1}{\alpha_g} \Big[ & \sum\limits_{u \in u^g}  \log  \mD_{ug} + \sum\limits_{j=1}^{M_g} \E_{\tilde{u} \sim P_{\gN}} \log (1 - \mD_{\tilde{u}g} ) \Big]
    \label{eqn:mi_loss}
\end{align}

where $\alpha_g = |g| + M_g$, $M_g$ is the number of negative users sampled for group $g$ and $\mD_{ug}$ is a shorthand for $\mD(\ve_u, \ve_g)$.
This objective maximizes MI between $\ve_u$ and $\ve_g$ based on the Jensen-Shannon divergence between the joint and the product of marginals~\cite{dgi}.

We employ a \textit{preference-biased} negative sampling distribution $P_{\gN} (\tilde{u} | g)$, which assigns higher likelihoods to \textit{non-member} users who have purchased the group items $\vx_{g}$.
These \textit{hard negative examples} encourage the discriminator to learn latent aspects shared by group members by contrasting against other users with similar individual item histories.
We define $P_{\gN} (\tilde{u} | g)$ as:
\begin{equation}
P_{\gN} (\tilde{u} | g) \propto  \eta \gI (\vx_{\tilde{u}}^T \cdot \vx_g > 0 \}) + (1 - \eta) \frac{1}{|\gU|}
\label{eqn:neg_sampler}
\end{equation}
where $\gI (\cdot)$ is an indicator function and $\eta$ controls the sampling bias.
We set $\eta=0.5$ across all our experiments.
In comparison to random negative sampling, our experiments indicate that preference-biased negative user sampling exhibits better discriminative abilities.

When $L_{MI}$ is trained jointly with the base recommender loss $L_R$ (equation~\ref{eqn:user_loss}), 
maximizing user-group MI enhances the quality of user and group representations computed by the encoder $f_{\textsc{enc}} (\cdot)$ and aggregator $f_{\textsc{agg}} (\cdot)$.
We now present our approach to overcome the limitations of the fixed regularizer $L_U$ (Section~\ref{sec:motivations}).

\subsection{\textbf{Contextual User Preference Weighting}}
\label{sec:context_weight}
In this section, we describe a contextual weighting strategy to identify and prioritize personal preferences of relevant group members, to overcome group interaction sparsity. 
We avoid degenerate solutions 
by varying the extent of regularization induced by each $\vx_u$ (for user $u$) across groups through group-specific relevance weights $w(u,g)$.
Contextual weighting accounts for user participation in diverse ephemeral groups with different levels of shared interests.

By maximizing user-group MI, the discriminator $\mD$ outputs scores $\mD(\ve_u, \ve_g)$ that quantify the contextual informativeness of each $(u,g)$ pair (higher scores for informative users).
Thus, we set the relevance weight $w(u,g)$ for group member $u \in u^g$ to be proportional to $\mD(\ve_u,\ve_g)$. 
Instead of regularizing the user representations $\mE$ with $\vx_u$ in each group ($L_U$ in eqn~\ref{eqn:user_loss}), we directly regularize the group representation $\ve_g$ with $\vx_u$ in proportion to $\mD(\ve_u, \ve_g)$ for each group member $u$. 
Direct optimization of $\ve_g$ (instead of $\ve_u$) results 
in more effective 
regularization, especially with sparse group activities.
We define the contextually weighted user-loss $L_{UG}$ as:

\begin{equation}
L_{UG} = - \sum\limits_{g \in \gG} \frac{1}{|\vx_g|} \sum\limits_{i \in \gI} \sum\limits_{u \in u^g} \mD(\ve_u, \ve_g) \; x_{ui} \log \pi_i (\ve_g)
\label{eqn:user_context_loss}
\end{equation}

where $L_{UG}$ effectively regularizes $\ve_g$ with the individual activities of group member $u$ with contextual weight $\mD(\ve_u,\ve_g)$.

The overall model objective of our framework~\name~includes three terms: $L_G$, $L_{UG}$, and $L_{MI}$, which is described in Section~\ref{sec:model_opt} in detail.
~\name~regularizes the latent representations computed by $f_{\textsc{enc}} (\cdot)$ and $f_{\textsc{agg}} (\cdot)$ through 
user-group MI maximization ($L_{MI}$) to contrastively capture group member associations; and contextual MI-guided weighting ($L_{UG}$) to prioritize individual preferences.

\subsection{Model Details}
\label{sec:model_details}
We now describe the architectural details of preference encoder $f_{\textsc{enc}}(\cdot)$, aggregator $f_{\textsc{agg}} (\cdot)$, discriminator $\mD$, and an alternative optimization approach to train our framework~\name.
\vspace{-4pt}
\subsubsection{\textbf{User Preference Encoder}}
To encode individual user preferences $\mX_U$ into preference embeddings $\mE$, we use a Multi-Layer Perceptron with two fully connected layers, defined by:
\[ \ve_u = f_{\textsc{enc}} (\vx_u) = \sigma (\mW_2^T (  \sigma  (\mW_1^T \vx_u + b_1 ) + b_2  ) \]
with learnable weight matrices $\mW_1 \in \sR^{|\gI| \times D}$ and $\mW_2 \in \sR^{D \times D}$, biases $b_1, b_2 \in \sR^D$, and $tanh(\cdot)$ activations for non-linearity $\sigma$.

\textbf{Pre-training:} We pre-train the weights and biases of the first encoder layer ($\mW_1, \vb_1$) on the user-item interaction matrix $\mX_U$ with user-loss $L_U$ (equation~\ref{eqn:user_loss}).
We use these pre-trained parameters to initialize the first layer of $f_{\textsc{enc}} (\cdot)$ before optimizing the overall objective of~\name.
Our ablation studies in Section~\ref{sec:ablation} indicate significant improvements owing to this initialization strategy.

\vspace{-4pt}
\subsubsection{\textbf{Group Preference Aggregators}}
\label{sec:group_agg}
We consider three preference aggregators \textsc{Maxpool}, \textsc{Meanpool} and, \textsc{Attention}, which are widely used 
in graph neural networks~\cite{gnn_powerful, graphsage, dysat} and have close ties to aggregators examined in prior work, \textit{i.e.}, \textsc{Maxpool} and \textsc{Meanpool} mirror the heuristics of maximum satisfaction~\cite{max_satisfaction} and averaging~\cite{average}, while attentions learn varying member contributions~\cite{agree, agr}.
We define the three preference aggregators below:

\begin{itemize}[leftmargin=*]
    \item \textbf{Maxpool}: The preference embedding of each member is passed through \textsc{MLP} layers, followed by element-wise max-pooling to aggregate group member representations, given by:
    \begin{equation*}
    \ve_g = \textsc{max}(\{ \sigma(\mW_{\textsc{agg}} \ve_{u} + b), \forall u \in u_g \})
    \end{equation*}
    where \textsc{max} denotes the element-wise max operator and $\sigma(\cdot)$ is a nonlinear activation.
    Intuitively, the MLP layers compute features for each member, and max-pooling over each of the computed features effectively captures different aspects of group members.
    \item \textbf{Meanpool}: We similarly apply an element-wise mean-pooling operation after the \textsc{MLP}, to compute group representation $\ve_g$ as:
    \begin{equation*}
        \ve_g = \textsc{mean}(\{ \sigma(\mW_{\textsc{agg}} \ve_{u} + b), \forall u \in u_g \}
    \end{equation*}
    \item \textbf{Attention}: To explicitly differentiate group members' roles, we employ neural attentions~\cite{attention} to compute a weighted sum of members' preference representations, where the weights are learned by an attention network, parameterized by a single \textsc{MLP} layer.
    \begin{equation*}
    \ve_g = \sum\limits_{u \in u_g} \alpha_u \mW_{\textsc{agg}} \ve_u \hspace{10 pt} \alpha_u = \frac{\exp (\vh^T \mW_{\textsc{agg}} \ve_u) }{\sum\limits_{u^{'} \in u_g} \exp(\vh^T \mW_{\textsc{agg}} \ve_{u^{'}}) }
    \end{equation*}
    where $\alpha_u$ indicates the contribution of a user $u$ towards the group decision. %
    This can be trivially extended to item-conditioned weighting~\cite{agree}, self-attention~\cite{wang2019group} and sub-attention networks~\cite{agr}.
\end{itemize}
\vspace{-4pt}
\subsubsection{\textbf{Discriminator Architecture}}
The discriminator architecture learns a scoring function to assign higher scores to observed $(u,g)$ pairs relative to negative examples, thus parameterizing group-specific relevance $w(u,g)$.
Similar to existing work~\cite{dgi}, we use a simple bilinear function to score user-group representation pairs.
\begin{equation}
    \mD (\ve_u, \ve_g) = \sigma (\ve_u^T \mW \ve_g)
\end{equation}
where $\mW$ is a learnable scoring matrix and $\sigma$ is the logistic sigmoid non-linearity function to convert raw scores into probabilities of $(\ve_u, \ve_g)$ being a positive example.
We leave investigation of further architectural variants for the discriminator $\mD$ to future work.

\vspace{-4pt}
\subsubsection{\textbf{Model Optimization}}
\label{sec:model_opt}
The overall objective of~\name~is composed of three terms, the group-loss $L_G$ (Equation~\ref{eqn:group_loss}), contextually weighted user-loss $L_{UG}$ (Equation~\ref{eqn:user_context_loss}), and MI maximization loss $L_{MI}$ (Equation~\ref{eqn:mi_loss}).
The combined objective is given by:

\begin{equation}
   L =  \hspace{-5pt} \underbrace{L_G}_{\text{Group Recommendation Loss}} \hspace{-25pt} + \hspace{-25pt} \overbrace{\lambda L_{UG}}^{\text{Contextually Weighted User Loss}} \hspace{-25pt}+ \hspace{-25pt} \underbrace{L_{MI}}_{\text{User-Group MI Maximization Loss}}
   \label{eqn:model_objective}
\end{equation}

We train~\name~using an alternating optimization schedule. In the first step, the discriminator $\mD$ is held constant, while optimizing the group recommender on $L_G + \lambda L_{UG}$.
The second step trains $\mD$ on $L_{MI}$, resulting in gradient updates for both parameters of $\mD$ as well as those of the encoder $f_{\textsc{enc}}(\cdot)$ and aggregator $f_{\textsc{agg}}(\cdot)$.

Thus, the discriminator $\mD$ only seeks to \textit{regularize} the model (\textit{i.e.,} encoder and aggregator) during training through loss terms $L_{MI}$ and $L_{UG}$.
During inference, we directly use the regularized encoder $f_{\textsc{enc}}(\cdot)$ and aggregator $f_{\textsc{agg}}(\cdot)$ to make group recommendations.

\section{Experiments}
\newcolumntype{K}[1]{>{\centering\arraybackslash}p{#1}}
\newcolumntype{R}[1]{>{\RaggedLeft\arraybackslash}p{#1}}

\begin{table}[t]
\centering

\begin{tabular}{@{}p{0.31\linewidth}R{0.12\linewidth}R{0.15\linewidth}R{0.14\linewidth}R{0.14\linewidth}@{}}

\toprule
\textbf{Dataset} &  \textbf{Yelp} & \textbf{Weeplaces} & \textbf{Gowalla}  & \textbf{Douban}\\
\midrule
\textbf{\# Users} &  7,037 & 8,643 & 25,405 & 23,032 \\
\textbf{\# Items} &  7,671 & 25,081 & 38,306 & 10,039\\
\textbf{\# Groups} &  10,214 & 22,733 & 44,565 & 58,983 \\
\textbf{\# U-I interactions} &  220,091 & 1,358,458 & 1,025,500 & 1,731,429 \\
\textbf{\# G-I Interactions} &  11,160 & 180,229 & 154,526 & 93,635 \\
\textbf{Avg. \# items/user} & 31.3 & 58.83 & 40.37 &  75.17\\
\textbf{Avg. \# items/group} & 1.09 & 2.95 & 3.47 &  1.59 \\
\textbf{Avg. group size} & 6.8 & 2.9 & 2.8 & 4.2 \\
\bottomrule
\end{tabular}
\caption{Summary statistics of four real-world datasets with ephemeral groups. Group-Item interactions are sparse: average number of interacted items per group < 3.5.}
\label{tab:dataset_stats}
\end{table}
In this section, we present an extensive quantitative and qualitative analysis of our model. We first introduce datasets, baselines, and experimental setup (Section~\ref{sec:datasets},~\ref{sec:baselines}, and~\ref{sec:setup}), followed by our main group recommendation results (Section~\ref{sec:main_results}).
In Section~\ref{sec:ablation}, we conduct an ablation study 
to understand our gains over the base recommender.
In Section~\ref{sec:group_char}, we study how key group characteristics (group \textit{size, coherence, and aggregate diversity}) impact recommendation results.
In Section~\ref{sec:mi_analysis}, we visualize the variation in discriminator scores assigned to group members, for different kinds of groups. Finally, we discuss limitations in Section~\ref{sec:limitations}.

\subsection{Datasets}
\label{sec:datasets}
First, we conduct experiments on large-scale POI (point-of-interest) recommendation datasets extracted from three location-based social networks.
Since the POI datasets do not contain explicit group interactions, we 
construct group interactions by jointly using the check-ins and social network information:
check-ins at the same POI within 15 minutes by an individual and her subset of friends in the social network together constitutes a single group interaction, while the remaining check-ins at the POI correspond to individual interactions.
We define the group recommendation task as recommending POIs to ephemeral groups of users.
The datasets were pre-processed to retain users and items with five or more check-ins each. We present dataset descriptions below:
\begin{itemize}[leftmargin=*]
    \item \textbf{Weeplaces}~\footnote{\url{https://www.yongliu.org/datasets/}}: we extract check-ins on POIs over all major cities in the United States, across various categories including Food, Nightlife, Outdoors, Entertainment and Travel.
    \item \textbf{Yelp}~\footnote{\url{https://www.yelp.com/dataset/challenge}}: we 
    filter the entire dataset to only include check-ins on restaurants located in the city of Los Angeles. 
    \item \textbf{Gowalla}~\cite{gowalla}: we use restaurant check-ins across all cities in the United States, in the time period upto June 2011.
\end{itemize}
    
Second, we evaluate venue recommendation on Douban, which is the largest online event-based social network in China.
\begin{itemize}[leftmargin=*]
    \item \textbf{Douban}~\cite{socialgroup}: users organize and participate in social events, where users attend events together in groups and items correspond to event venues.
    During pre-processing, 
    we filter out users and venues with less than 10 interactions each.
\end{itemize}
Groups across all datasets are ephemeral since group interactions are sparse (average number of items per group $< 3.5$ in Table~\ref{tab:dataset_stats})

\newcommand*{\factor}{0.031}
\begin{table*}[ht]
\centering
\small
\begin{tabular}{@{}p{0.16\linewidth}@{\hspace{13pt}}
K{\factor\linewidth}K{\factor\linewidth}K{\factor\linewidth}K{\factor\linewidth}@{\hspace{13pt}}
K{\factor\linewidth}K{\factor\linewidth}K{\factor\linewidth}K{\factor\linewidth}@{\hspace{13pt}}
K{\factor\linewidth}K{\factor\linewidth}K{\factor\linewidth}K{\factor\linewidth}@{\hspace{13pt}}
K{\factor\linewidth}K{\factor\linewidth}K{\factor\linewidth}K{\factor\linewidth}K{\factor\linewidth}@{}} \\
\toprule
\multirow{1}{*}{\textbf{Dataset}} &  \multicolumn{4}{c}{\textbf{Yelp (LA)}} & \multicolumn{4}{c}{\textbf{Weeplaces}}  & \multicolumn{4}{c}{\textbf{Gowalla}} & \multicolumn{4}{c}{\textbf{Douban}} \\
\multirow{1}{*}{\textbf{Metric}} & \textbf{N@20} & \textbf{N@50} & \textbf{R@20} & \textbf{R@50} & \textbf{N@20} & \textbf{N@50} & \textbf{R@20} & \textbf{R@50}  & \textbf{N@20}& \textbf{N@50} &
\textbf{R@20} & \textbf{R@50} & \textbf{N@20}& \textbf{N@50} & \textbf{R@20} & \textbf{R@50}\\
\midrule
\multicolumn{17}{c}{\textbf{Predefined Score Aggregators}} \\
\midrule
\textbf{Popularity~\cite{popularity}}  & 0.000  & 0.000  & 0.001  & 0.001  & 0.063  & 0.074  & 0.126  & 0.176  & 0.075  & 0.088  & 0.143  & 0.203  & 0.003  & 0.005  & 0.009  & 0.018  \\
\textbf{VAE-CF + AVG~\cite{vae-cf,average}}  & 0.142  & 0.179  & 0.322  & 0.513  & 0.273  & 0.313  & 0.502  & 0.666  & 0.318  & 0.362  & 0.580  & 0.758  & 0.179  & 0.217  & 0.381  & 0.558  \\
\textbf{VAE-CF + LM~\cite{vae-cf,least_misery}}  & 0.097  & 0.120  & 0.198  & 0.316  & 0.277  & 0.311  & 0.498  & 0.640  & 0.375  & 0.409  & 0.610  & 0.750  & 0.221  & 0.252  & 0.414  & 0.555  \\
\textbf{VAE-CF + MAX~\cite{vae-cf,max_satisfaction}}  & 0.099  & 0.133  & 0.231  & 0.401  & 0.229  & 0.270  & 0.431  & 0.604  & 0.267  & 0.316  & 0.498  & 0.702  & 0.156  & 0.194  & 0.339  & 0.517  \\
\textbf{VAE-CF + RD~\cite{vae-cf,rd}}  & 0.143  & 0.181  & 0.321  & 0.513  & 0.239  & 0.279  & 0.466  & 0.634  & 0.294  & 0.339  & 0.543  & 0.723  & 0.178  & 0.216  & 0.379  & 0.557  \\
\midrule
\multicolumn{17}{c}{\textbf{Data-driven Preference Aggregators}}\\
\midrule
\textbf{COM~\cite{com}}  & 0.143  & 0.154  & 0.232  & 0.286  & 0.329  & 0.348  & 0.472  & 0.557  & 0.223  & 0.234  & 0.326  & 0.365  & 0.283  & 0.288  & 0.417  & 0.436  \\
\textbf{Crowdrec~\cite{crowdrec}}  & 0.082  & 0.101  & 0.217  & 0.315  & 0.353  & 0.370  & 0.534  & 0.609  & 0.325  & 0.338  & 0.489  & 0.548  & 0.121  & 0.188  & 0.375  & 0.681  \\
\textbf{AGREE~\cite{agree}}  & 0.123  & 0.168  & 0.332  & 0.545  & 0.242  & 0.292  & 0.484  & 0.711  & 0.160  & 0.223  & 0.351  & 0.605  & 0.126  & 0.173  & 0.310  & 0.536  \\
\textbf{MoSAN~\cite{agr}}  & 0.470  & 0.494  & 0.757  & \textbf{0.875}  & 0.287  & 0.334  & 0.548  & 0.738  & 0.323  & 0.372  & 0.584  & 0.779  & 0.193  & 0.239  & 0.424  & 0.639  \\
\midrule
\multicolumn{17}{c}{\textbf{Group Information Maximization Recommenders (GroupIM)}}\\
\midrule
\textbf{{GroupIM-Maxpool} } & 0.488  & 0.501 & 0.676  & 0.769   & 0.479  & 0.505  & 0.676  & 0.776  & 0.433  & 0.463  & 0.628  & 0.747  & 0.291 & 0.313   & 0.524  & 0.637   \\
\textbf{{GroupIM-Meanpool}}  & 0.629  & 0.637  & 0.778  & 0.846  & 0.518  & 0.543  & 0.706  & 0.804  & 0.476  & 0.504  & 0.682  & 0.788  & 0.323  & 0.351  & 0.569  & 0.709  \\
\textbf{{GroupIM-Attention}}  & \textbf{0.633}  & \textbf{0.647}  &\textbf{0.782}  & 0.851   & \textbf{0.521}  & \textbf{0.546}  & \textbf{0.716}  & \textbf{0.813}  & \textbf{0.477}  & \textbf{0.505}  & \textbf{0.686}  & \textbf{0.796} &  \textbf{0.325}  & \textbf{0.356}  & \textbf{0.575}  & \textbf{0.714}   \\
\bottomrule
\end{tabular}
\caption{Group recommendation results on four datasets, R@K and N@K denote the \textsc{Recall}@K and \textsc{NDCG}@K metrics at $K = 20$ and $50$.
The~\name~variants indicate \textsc{maxpool}, \textsc{meanpool}, and \textsc{attention} as preference aggregators in our MI maximization framework.
~\name~achieves significant gains of 31 to 62\% NDCG@20 and 3 to 28\% \textsc{Recall}@20 over competing group recommenders. 
Notice that \textsc{meanpool} and \textsc{attention} variants achieve comparable performance across all datasets.
}
\vspace{2pt}
\label{tab:main_results}
\end{table*}

\subsection{Baselines}
\label{sec:baselines}
We compare our framework against state-of-the-art baselines that broadly fall into two categories: score aggregation methods with predefined aggregators, and data-driven preference aggregators.

\begin{itemize}[leftmargin=*]
    \item \textbf{Popularity}~\cite{popularity}: recommends items based on item popularity, which is measured by its interaction count in the training set. 
    \item \textbf{User-based CF + Score Aggregation}: We utilize a 
    state-of-the-art neural recommendation model VAE-CF~\cite{vae-cf}, followed by score aggregation via:
averaging (AVG), least-misery (LM), maximum satisfaction (MAX), and relevance-disagreement (RD). %
    \item \textbf{COM}~\cite{com}: a probabilistic generative model that %
    considers
    group members' individual preferences and topic-dependent influence.
    \item \textbf{CrowdRec}~\cite{crowdrec}: a generative model that extends COM through item-specific latent variables capturing their global popularity.
    \item \textbf{MoSAN}~\cite{agr}: a neural group recommender that employs a collection of sub-attentional networks to model group member interactions. Since MoSAN originally ignores individual activities $\mX_U$, we include $\mX_U$ into $\mX_G$ as pseudo-groups with single users.
    \item \textbf{AGREE}~\cite{agree}: a neural group recommender that utilizes attentional preference aggregation to compute item-specific group member weights, for joint training over individual and group activities.
\end{itemize}

\begin{figure*}[t]
    \vspace{-13pt}
    \centering
    \includegraphics[width=0.972\linewidth]{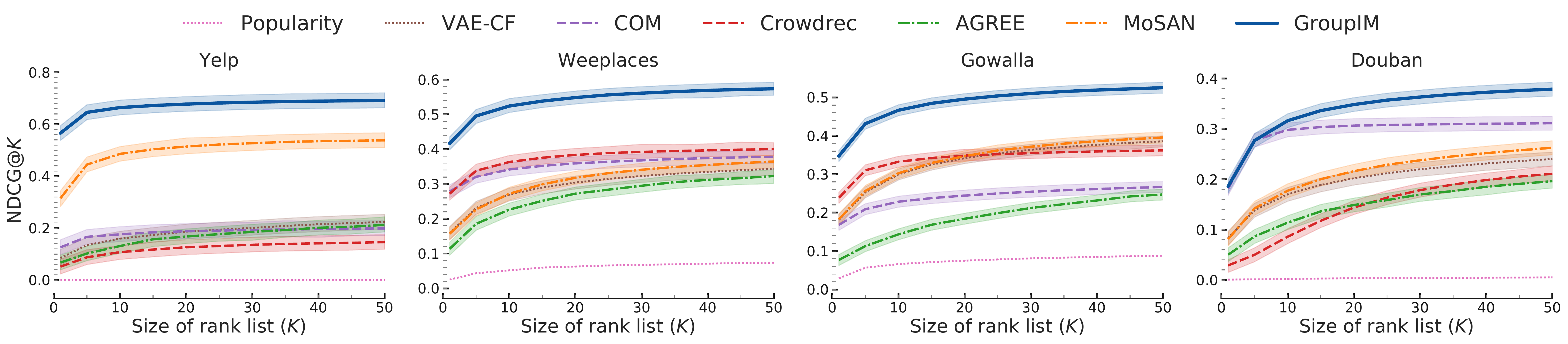}
    \caption{\textsc{NDCG}@K across size of rank list $K$. Variance bands indicate 95\% confidence intervals over 10 random runs.
    Existing methods underperform since they either disregard member roles (VAE-CF variants) or overfit to the sparse group activities.
    ~\name~contextually identifies informative members and regularizes their representations, to show strong gains.}
    \label{fig:ndcg}
\end{figure*}

We tested~\textbf{\name}~by substituting three preference aggregators, \textsc{Maxpool}, \textsc{Meanpool}, and \textsc{Attention} (Section~\ref{sec:group_agg}). 
All experiments were conducted on a single Nvidia Tesla V100 GPU with \textit{PyTorch}~\cite{pytorch} implementations on the Linux platform. Our implementation of~\textbf{\name}~and datasets are publicly available\footnote{https://github.com/CrowdDynamicsLab/GroupIM}.

\subsection{Experimental Setup}
\label{sec:setup}
We randomly split the set of all groups into training (70\%), validation (10\%), and test (20\%) sets, while utilizing the individual interactions of all users for training.
Note that each group appears only in one of the three sets.
The test set contains \textit{strict ephemeral groups} (\textit{i.e.}, a specific combination of users) that do not occur in the training set. Thus, we train on ephemeral groups and test on strict ephemeral groups.
We use \textsc{NDCG@K} and \textsc{Recall@K} as evaluation metrics. %

We tune the latent dimension 
in the range $\{ 32, 64, 128\}$ 
and other baseline hyper-parameters in ranges centered at author-provided values. %
In GroupIM, we use two fully connected layers of size 64 each
in $f_{\textsc{enc}} (\cdot)$
and tune $\lambda$ in the range $\{2^{-4}, 2^{-3}, \dots, 2^6\}$.
We use 5 negatives for each true user-group pair to train the discriminator.

\subsection{Experimental Results}
\label{sec:main_results}
We note the following key observations from our experimental results comparing~\textbf{\name}~with its three aggregator variants, against competing baselines on group recommendation (Table~\ref{tab:main_results}).

First, heuristic score aggregation with neural recommenders (\textit{i.e.},~\textbf{VAE-CF}) performs comparable to (and often beats) 
probabilistic models 
(\textbf{COM, Crowdrec}).
Neural methods with multiple non-linear transformations, are expressive enough to identify latent groups of similar users just from their individual interactions.

Second, there is no clear winner among the different pre-defined score aggregation strategies, \textit{e.g.}, \textbf{VAE-CF + LM} (least misery) outperforms the rest on Gowalla and Douban, while \textbf{VAE-CF + LM} (averaging) is superior on Yelp and Weeplaces. This empirically validates the
non-existence of a single optimal strategy for all datasets.

Third,~\textbf{MoSAN}~\cite{agr} outperforms both probabilistic models and fixed score aggregators on most datasets.
~\textbf{MoSAN}~achieves better results owing to the expressive power of neuural preference aggregators (such as sub-attention networks) to capture group member interactions, albeit not explicitly differentiating personal and group activities.
Notice that naive joint training over personal and group activities via static regularization (as in \textbf{AGREE}~\cite{agree}) results in poor performance due to sparsity in group interactions.
Static regularizers on $\mX_U$ cannot distinguish the role of users across groups, resulting in models that lack generalisation to ephemeral groups.

~\textbf{\name}~variants outperform baselines significantly, with \textsc{attention} achieving overall best results.
In contrast to neural methods (\textit{i.e.}, \textbf{MoSAN} and \textbf{AGREE}), 
~\textbf{\name}~regularizes the latent representations by contextually weighting the personal preferences of informative members, thus effectively tackling group interaction sparsity.
The \textsc{maxpool} variant is noticeably inferior, due to the higher sensitivity of \textit{max} operation to outlier group members.

Note that \textsc{Meanpool} performs comparably to \textsc{attention}. This is because in~\textbf{\name}, the discriminator $\mD$ does the heavy-lifting of contextually differentiating the role of users across groups to effectively regularize the encoder $f_{\textsc{enc}} (\cdot)$ and aggregator $f_{\textsc{agg}} (\cdot)$ modules.
If $f_{\textsc{enc}} (\cdot)$ and $\mD$ are expressive enough, efficient \textsc{meanpool} aggregation can achieve near state-of-the-art results (Table~\ref{tab:main_results})

An important implication is the \textit{reduced inference complexity} of our model, \textit{i.e.}, once trained using our MI maximizing framework, simple aggregators (such as \text{meanpool}) suffice to achieve state-of-the-art performance.
This is especially significant, considering that our closest baseline \textbf{MoSAN}~\cite{agr} utilizes sub-attentional preference aggregation networks that scale quadratically with group size.

We compare the variation in \textsc{NDCG} scores with size of rank list in figure~\ref{fig:ndcg}. We only depict the best aggregator for \textbf{VAE-CF}.
~\textbf{\name} consistently generates more \textit{precise} recommendations across all datasets. We observe smaller gains in Douban, where the user-item interactions exhibit substantial correlation with corresponding group activities.~\textbf{\name} achieves significant gains in characterizing diverse groups, evidenced by our results in section~\ref{sec:group_char}.

\renewcommand*{\factor}{0.088}
\begin{table}[t]
\vspace{-5pt}
\centering
\small
\begin{tabular}{@{}p{0.47\linewidth}K{\factor\linewidth}K{\factor\linewidth}
K{\factor\linewidth}@{\hspace{13pt}}K{\factor\linewidth}K{\factor\linewidth}K{\factor\linewidth}@{}} \\
\toprule
{\textbf{Dataset}} &  \multicolumn{2}{c}{\textbf{Weeplaces} } & \multicolumn{2}{c}{\textbf{Gowalla}} \\
\textbf{Metric} & \textbf{N@50} & \textbf{R@50} & \textbf{N@50} &  \textbf{R@50} \\
\midrule
\multicolumn{5}{c}{\textbf{Base Group Recommender Variants}} \\
\midrule
(1) \textbf{Base} ($L_G$) & 0.420 &  0.621  &  0.369 & 0.572  \\
(2) \textbf{Base} ($L_G  + \lambda L_U$) & 0.427 & 0.653 & 0.401  & 0.647 \\
\midrule
\multicolumn{5}{c}{\textbf{\name~Variants}} \\
\midrule
(3) \textbf{\name} ($L_G + L_{MI}$) & 0.431 & 0.646 & 0.391 & 0.625 \\
(4) \textbf{\name} (Uniform weights)  &  0.441  &  0.723 & 0.418 &  0.721 \\
(5) \textbf{\name} (Cosine similarity) & 0.488 & 0.757 &  0.445 & 0.739 \\
(6) \textbf{\name} (No pre-training) & 0.524 & 0.773 &  0.472 & 0.753 \\
(7) \textbf{\name} ($L_G+ \lambda L_{UG} + L_{MI}$)  & \textbf{0.543} & \textbf{0.804} & \textbf{0.505} & \textbf{0.796}  \\
\bottomrule
\end{tabular}
\caption{~\name~ablation study ($\textsc{NDCG}$ and $\textsc{Recall}$ at $K = 50$).
Contrastive representation learning (row 3) improves the base recommender (row 1), but is substantially more effective with group-adaptive preference weighting (row 7).}
\label{tab:ablation_results}
\end{table}

\subsection{Model Analysis}
\label{sec:ablation}
In this section, we present an ablation study to analyze several variants of~\textbf{\name}, guided by our motivations (Section~\ref{sec:motivations}).
In our experiments, we choose \textsc{attention} as the aggregator due to its consistently high performance.
We conduct studies on Weeplaces and Gowalla to report \textsc{NDCG}@50 and \textsc{Recall}@50 in Table~\ref{tab:ablation_results}.

First, we examine the base group recommender $\mR$ (Section~\ref{sec:base_recommender}) which does not utilize MI maximization for model training (Table~\ref{tab:ablation_results}).

\vspace{-4pt}
\subsubsection*{\textbf{Base Group Recommender.}} We examine two variants below:\\
(1) We train $\mR$ on just group activities $\mX_G$ with loss $L_G$ (equation~\ref{eqn:group_loss}).\\
(2) We train $\mR$ jointly on individual $\mX_U$ and group $\mX_G$ activities with static regularization on $\mX_U$ using joint loss $L_R$ (Equation~\ref{eqn:user_loss}).

In comparison to similar neural aggregator \textbf{MoSAN}, our base recommender $\mR$ is stronger on \textsc{NDCG} but inferior on \textsc{Recall}.
The difference is likely due to the multinomial likelihood used to train $\mR$, in contrast to the ranking loss in \textbf{MoSAN}.
Static regularization via $\mX_U$ (row 1) results in higher gains for Gowalla (richer user-item interactions) with relatively larger margins for \textsc{Recall} than \textsc{NDCG}.

Next, we examine model variants of~\textbf{\name}~in two parts:
\vspace{-4pt}
\subsubsection*{\textbf{\name: Contrastive Representation Learning.}} We analyze the benefits derived by just training the contrastive discriminator $\mD$ to capture group member associations, \textit{i.e.}, we define a model variant (row 3) to optimize just $L_G + L_{MI}$, without the $L_{UG}$ term.
Direct MI maximization (row 3)
improves over the base recommender $\mR$ (row 1), validating the benefits of contrastive regularization, however still suffers from lack of user preference prioritization.
\vspace{-4pt}

\subsubsection*{\textbf{\name: Group-adaptive Preference Prioritization}}
We analyze the utility of data-driven contextual weighting (via user-group MI), 
by examining two alternate \textit{fixed} strategies to define $w(u,g)$:\\ %
(4) \textbf{\textit{Uniform weights}}: We assign the same weight $w(u,g) = 1$ for each group member $u$ in group $g$, when optimizing $L_{UG}$. \\
(5) \textbf{\textit{Cosine similarity}}: To model user-group correlation, we set the weight $w(u,g)$ as the cosine similarity between $\vx_u$ and $\vx_g$.

From table~\ref{tab:ablation_results} (rows 4 and 5), the uniform weights variant of loss $L_{UG}$ (row 4) surpasses the statically regularized model (row 2), due to more direct feedback from $\mX_U$ to group embedding $\ve_g$ during model training.
Cosine similarity (row 5) achieves stronger gains owing to more accurate correlation-guided user weighting across groups.
Our model~\textbf{\name}~(row 7) has strong gains over the fixed weighting strategies as a result of its regularization strategy to contextually identify informative members across groups. 
\vspace{-4pt}
\subsubsection*{\textbf{\name: Pre-training $f_{\textsc{enc}} (\cdot)$ on $\mX_U$}}
We depict model performance without pre-training (random initializations) in row 6.
Our model (row 7) achieves noticeable gains; pre-training identifies good model initialization points for better convergence.

\vspace{-4pt}
\subsection{Impact of Group Characteristics}
\label{sec:group_char}
In this section, we examine our results to understand the reason for~\textbf{\name}'s gains. We study ephemeral groups along three facets: \textit{group size}; \textit{group coherence}; and \textit{group aggregate diversity}.

\begin{figure}[t]
    \centering
    \includegraphics[width=\linewidth]{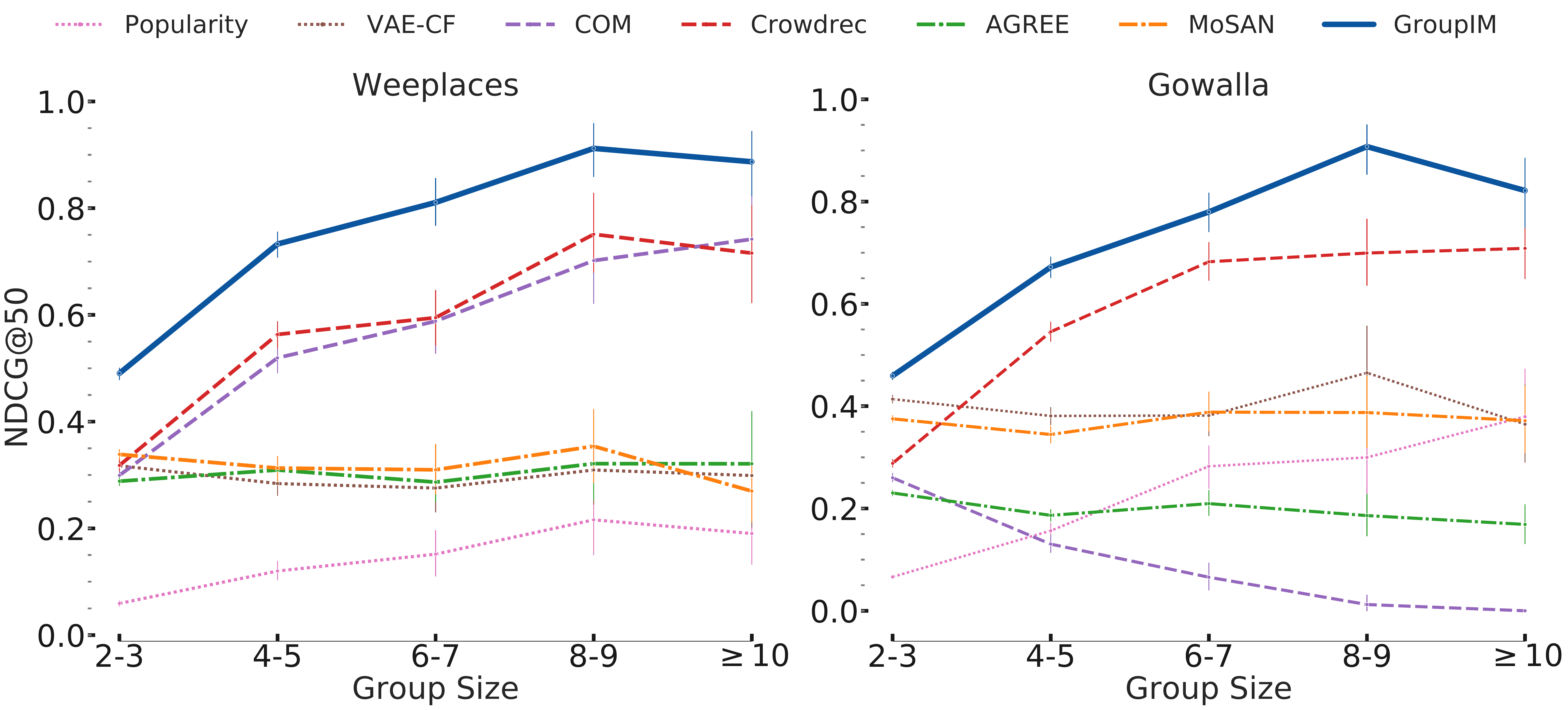}
    \caption{Performance (NDCG@50), across group size ranges.~\name~has larger gains for larger groups due to accurate user associations learnt via MI maximization.}
    \label{fig:group_size}
\end{figure}

\vspace{-4pt}
\subsubsection{\textbf{Group Size}}
We classify test groups into bins based on five levels of group size (2-3, 4-5, 6-7, 8-9, and  $\geq$10).
Figure~\ref{fig:group_size} depicts the variation in \textsc{NDCG}@50 scores on Weeplaces and Gowalla datasets.

We make three key observations:  methods that explicitly distinguish individual and group activities (such as \textbf{COM, CrowdRec, GroupIM}), exhibit distinctive trends \textit{wrt} group size. In contrast, \textbf{MoSAN}~\cite{agr} and \textbf{AGREE}~\cite{agree}, which either uniformly mix both behaviors or apply static regularizers, show no noticeable variation; %
Performance generally increases with group size. 
Although test groups are previously unseen, for larger groups, 
subsets of inter-user interactions are more likely to be seen during training,
thus resulting in better performance; 
~\textbf{\name}~achieves higher (or steady) gains for groups of larger sizes owing to its more accurate prioritization of personal preferences for each member, \textit{e.g.}, ~\textbf{\name}~clearly has stronger gains for groups of sizes 8-9 and $\geq10$ in Gowalla.
\vspace{-4pt}
\subsubsection{\textbf{Group Coherence}}
\begin{figure}[t]
    \centering
    \includegraphics[width=\linewidth]{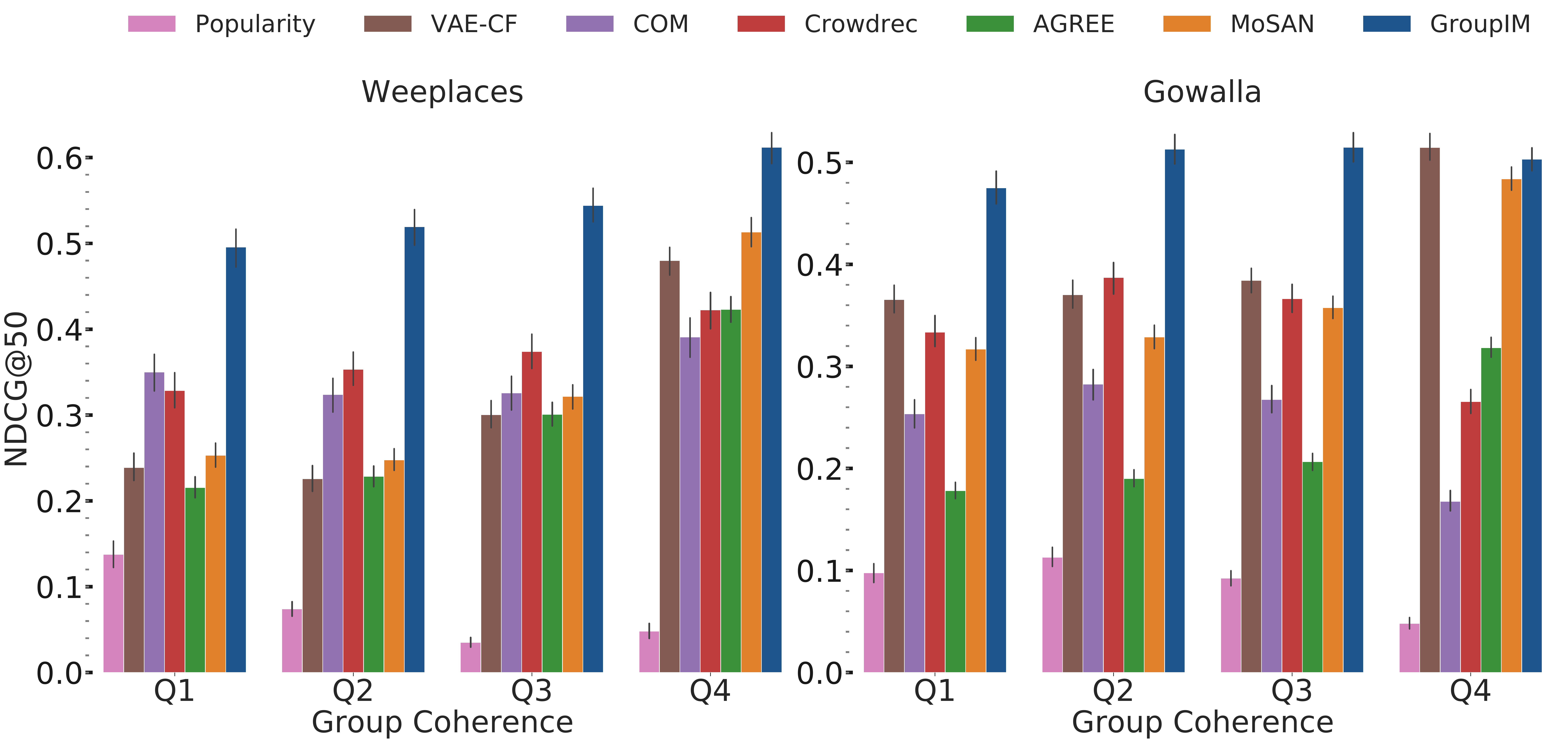}
    \caption{Performance (NDCG@50), across group coherence quartiles (Q1: lowest, Q4: highest).~\name~has larger gains in Q1 \& Q2 (low group coherence).}
    \label{fig:group_coherence}
\end{figure}

We define \textit{group coherence} as the mean pair-wise correlation of personal activities ($\vx_u$) of group members, \textit{i.e.}, if a group has users who frequently co-purchase items, it receives greater coherence.
We separate test groups into four quartiles by their coherence scores.
Figure~\ref{fig:group_coherence} depicts \textsc{NDCG}@50
for groups under each quartile (Q1 - Lower values), on Weeplaces and Gowalla.

~\textbf{\name}~has stronger gains for groups with low coherence (quartiles Q1 and Q2), which empirically validates the efficacy of contextual user preference weighting in regularizing the encoder and aggregator, for groups with dissimilar member preferences.
\vspace{-4pt}

\begin{figure}[t]
    \centering
    \includegraphics[width=\linewidth]{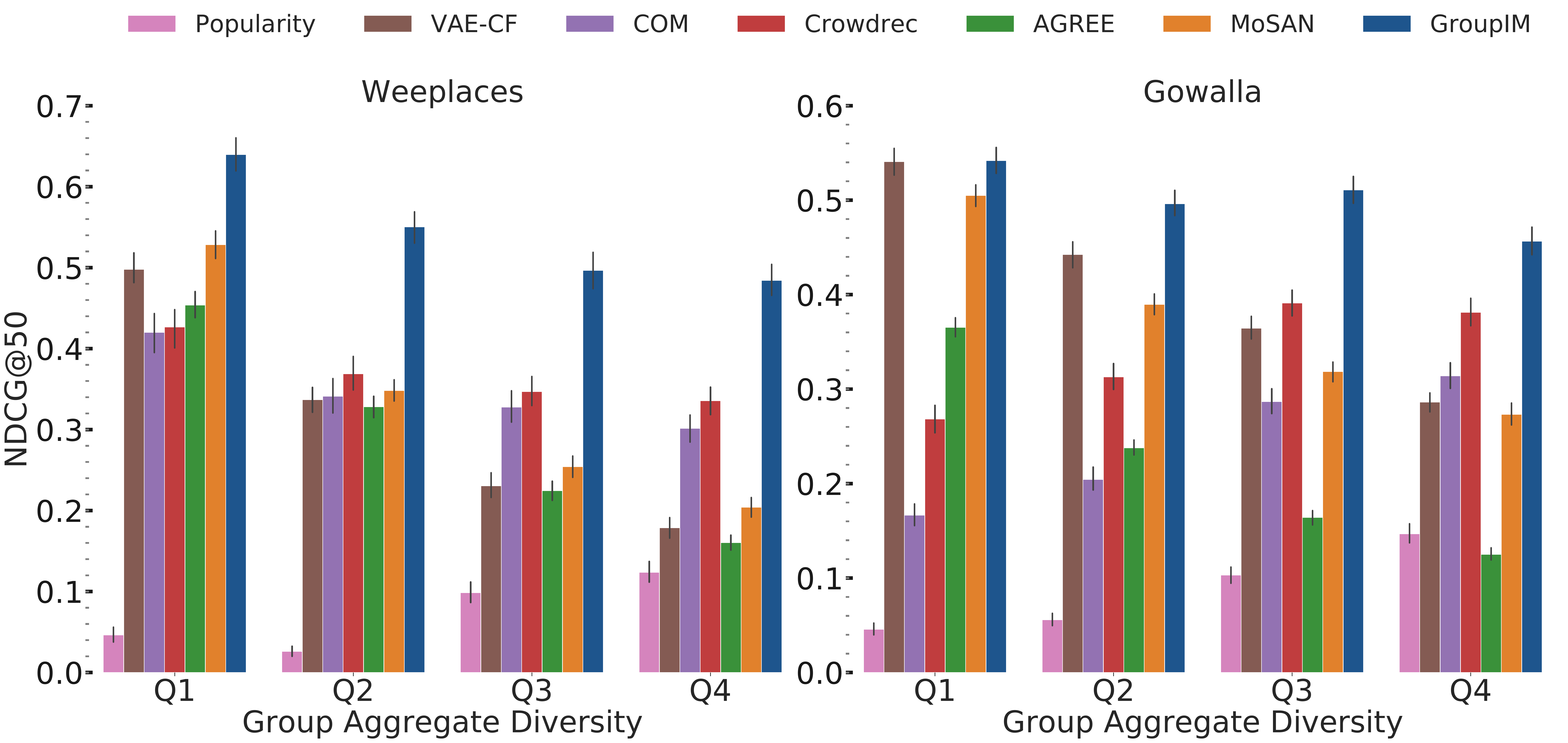}
    \caption{Performance (NDCG@50), across group aggregate diversity quartiles (Q1: lowest, Q4: highest).~\name~has larger gains in Q3 \& Q4 (high diversity).}
    \label{fig:group_coverage}
\end{figure}
\subsubsection{\textbf{Group Aggregate Diversity}}
We adapt the classical \textit{aggregate diversity} metric~\cite{diversity_metrics} to define \textit{group aggregate diversity} as the total number of distinct items interacted across all group members, 
\textit{i.e.}, if the set of all purchases of group members covers a wider range of items, then the group has higher aggregate diversity. 
We report NDCG@50 across aggregate diversity quartiles in figure~\ref{fig:group_coverage}.

Model performance typically decays (and stabilizes), with increase in aggregate diversity.
Diverse groups with large candidate item sets, pose an information overload for group recommenders, leading to worse results. 
Contextual prioritization with contrastive learning, benefits diverse groups, as evidenced by the higher relative gains of~\textbf{\name}~ for diverse groups (quartiles Q3 and Q4).

\begin{figure*}[t]
    \centering
    \includegraphics[width=\linewidth]{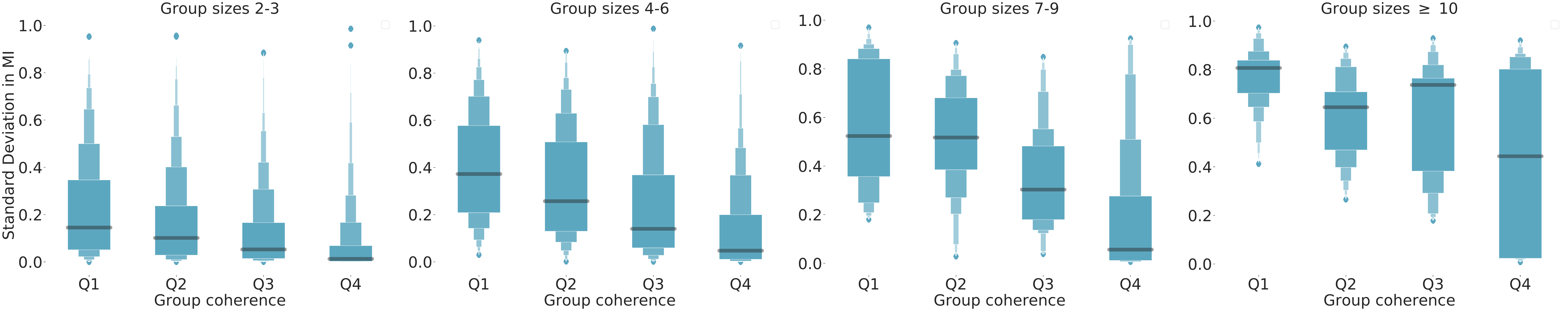}
    \caption{\textit{MI variation} (std. deviation in discriminator scores over members) per group coherence quartile across group sizes.
    For groups of a given size, as coherence increases, \textit{MI variation} decreases.
    As groups increase in size, \textit{MI variation} increases.}
    \label{fig:discriminator}
\end{figure*}

\subsection{Qualitative MI Discriminator Analysis}
\label{sec:mi_analysis}
We examine the contextual weights $w(u,g)$ estimated by \textbf{\name} over test ephemeral groups, across group size and coherence.

We divide groups into four bins based on group sizes (2-3, 4-6, 7-9, and $\geq10$), and partition them into quartiles based on group coherence within each bin.
To analyze the variation in \textit{contextual informativeness} across group members, 
we compute \textit{MI variation} as the standard deviation of scores given by $\mD$ over group members.
Figure~\ref{fig:discriminator} depicts letter-value plots of \textit{MI variation} for groups in corresponding coherence quartiles across group sizes on Weeplaces.

\textit{MI variation} increases with group size, since larger groups often comprise users with divergent roles and interests. 
Thus, the discriminator generalizes to unseen groups, to discern and estimate markedly different relevance scores for each group member.
To further examine the intuition conveyed by the scores, we compare \textit{MI variation} across group coherence quartiles within each size-range.

\textit{MI variation} is negatively correlated with group coherence for groups of similar sizes, \textit{e.g.}, \textit{MI variation} is consistently higher for groups with low coherence (quartiles Q1 and Q2).
For highly coherent groups (quartile Q4), $\mD$ assigns comparable scores across all members, which is consistent with our intuitions and earlier results on the efficacy of simple averaging strategies for such groups.

\label{sec:sensitivity}
We also analyze parameter sensitivity to user-preference weight $\lambda$. Low $\lambda$ values result in overfitting to the group activities $\mX_G$, while larger values result in degenerate solutions that lack group distinctions (plot excluded for the sake of brevity).

\subsection{Limitations}
\label{sec:limitations}
We identify two limitations of our work. Despite \textit{learning} to contextually prioritize users' preferences across groups, $\lambda$ controls the overall strength of preference regularization.
Since optimal $\lambda$ varies across datasets and applications, we plan to explore meta-learning approaches to eliminate such hyper-parameters~\cite{meta-reweight}.

\textbf{\name}~relies on user-group MI estimation to contextually identify informative members, which might become challenging when users have sparse individual interaction histories.
In such a scenario, side information (\textit{e.g.}, social network of users), or contextual factors (\textit{e.g.}, location, interaction time)~\cite{context} can prove effective.

\section{Conclusion}
This paper introduces a recommender architecture-agnostic framework~\name~that integrates arbitrary neural preference encoders and aggregators for ephemeral group recommendation.
To overcome group interaction sparsity,~\name~regularizes the user-group representation space by maximizing user-group MI to contrastively capture preference covariance among group members. %
Unlike prior work that incorporate individual preferences through static regularizers, we
dynamically prioritize the preferences of informative members through MI-guided contextual preference weighting.
Our extensive experiments on four real-world datasets show significant gains for~\name~over state-of-the-art methods.

\bibliographystyle{ACM-Reference-Format}
\bibliography{main}
\appendix

\end{document}